\documentclass[aps,prb,twocolumn,floatfix,amsmath,amssymb]{revtex4}
\usepackage{graphicx}
\usepackage{dcolumn}
\usepackage{bm}
\usepackage{epsfig,psfrag,amsmath,amssymb,float}
\input{epsf}
\begin{document}

\title{Asymmetric electron-phonon interactions 
in the three-band Peierls-Hubbard model}
\author{Z.~B.~Huang, W.~Hanke, and E.~Arrigoni}
\affiliation{Institut f\"ur Theoretische Physik, 
Universit\"at W\"urzburg, am Hubland,
97074 W\"urzburg, Germany}

\date{\today}

\begin{abstract}
Using the Quantum Monte Carlo (QMC) technique within frozen-phonon,
we studied the effects of the half-breathing O$(\pi,0)$ phonon mode on 
the ground-state properties of the three-band Peierls-Hubbard model. 
Our simulations are performed for both ionic and covalent electron-phonon
couplings. The effects of lattice displacements on the ground-state 
energies and charge fluctuations are similar in magnitude for both hole- and
electron-doped cases. However, the effects of lattice displacements on the 
magnetic properties are rather different. In the hole-doped case, the 
normalized next-nearest-neighbor Cu-Cu spin correlations are dramatically 
modified by both ionic and covalent electron-phonon couplings. 
On the other hand, in the electron-doped case, much smaller effects are 
observed. The distinct spin-phonon couplings, in conjunction with the 
spin-bag picture of the quasiparticle, could explain a strong mass 
renormalization effect in the p-type cuprates and a weaker effect 
in the n-type cuprates.
\end{abstract}

\maketitle

Recently, a large number of experimental observations in the superconducting 
cuprates suggest that the electron-phonon (el-ph) coupling in these systems
is strong and -in conjunction with strong electronic correlations, 
may play a role for the superconductivity mechanism.
In particular, angle resolved photoemission spectroscopy (ARPES) data 
indicate a mass renormalization (``kink")
in the electronic dispersion of the p-type high-T$_c$ cuprates near a 
characteristic energy scale ($\approx 70$ meV)~\cite{lanzara,shen}, which is 
possibly caused by coupling of quasiparticle to some phonon modes.
Moreover, ARPES data~\cite{armitage} also show that
this mass renormalization is weak in the n-type 
Nd$_{1.85}$Ce$_{0.15}$CuO$_4$, indicating that the el-ph coupling 
in the p-type materials is much stronger than in the n-type ones.
On the other hand, both inelastic neutron scattering~\cite{queeney,pint,kang}
and inelastic x-ray scattering~\cite{astuto} found that the anomalous
softening of the Cu-O bond-stretching oxygen mode is present in
both p- and n-type materials, suggesting that although there exist
some differences, the el-ph coupling is strong in both kinds of materials.

In order to understand this asymmetry of el-ph coupling in
p- and n-type cuprates, we have studied
the three-band Peierls-Hubbard model in the physically interesting
parameter regime by employing the constrained-path Monte Carlo 
(CPMC) method~\cite{zhang,carlson}.  In the CPMC method,
the ground-state wave function $|\Psi_0\rangle$ is projected from 
a known trial wave function $|\Psi_T\rangle$ and then 
the physical quantities are calculated in this ground state.
We find that in both hole- and electron-doped cases, the effects of 
lattice displacements on the ground-state energies and charge 
fluctuations are similar. However, an essential difference between hole- 
and electron-doped systems is that the response of spin degree of 
freedom to phonons is strong in the former case, but rather weak 
for the latter.

In the hole representation the three-band Peierls-Hubbard 
model~\cite{yone,yu} has the Hamiltonian,
\begin{eqnarray}
H&=&\sum_{\langle i,j\rangle\sigma}t_{pd}^{ij}(\{u_{j}\})(d_{i\sigma}^{\dagger}p_
{j\sigma}+h.c.)+\sum_{\langle j,k\rangle\sigma}t_{pp}^{jk}(p_{j\sigma}^{\dagger}
p_{k\sigma}+h.c.)
\nonumber\\
&&+\epsilon_{p}\sum_{j\sigma}n_{j\sigma}^{p}
+\sum_{i\sigma}\epsilon_{d}^{i}(\{u_{j}\})n_{i\sigma}^{d}
\nonumber\\
&&+U_{d}\sum_{i}n_{i\uparrow}^{d}n_{i\downarrow}^{d}+
U_{p}\sum_{j}n_{j\uparrow}^{p}n_{j\downarrow}^{p}+V_{pd}\sum_{\langle i,j\rangle
}n_{i}^{d}n_{j}^{p}.
\end{eqnarray}
Here the operator $d_{i\sigma}^{\dagger}$ creates a hole at a Cu
$3d_{x^2-y^2}$ orbital and $p_{j\sigma}^{\dagger}$ creates a hole in an O
$2p_x$ or $2p_y$ orbital. $U_d$ and $U_p$ are the Coulomb energies 
at the Cu and O sites, respectively.
$V_{pd}$ denotes the nearest-neighbor (nn) Coulomb repulsion. 
For the el-ph coupling, we consider that the nn Cu-O 
hybridization is modified by the O-ion displacement $u_j$ as 
$t_{pd}^{ij} = \phi_{ij} (t_{pd}\pm\alpha u_{j})$, where the $+ (-)$ applies 
if the bond shrinks (stretches) with positive $u_{j}$.
The phase factor $\phi_{ij}$ takes a minus sign for $j = i + \hat{x}/2 $ and
$j = i - \hat{y}/2 $. The Cu-site energy is assumed to be modulated by
the nn O-ion displacements $u_j$ as 
$\epsilon_{d}^{i}(\{u_{j}\})=\epsilon_{d}+\beta\sum_{j}\pm u_{j}$.
The other electronic matrix elements are the O-O hybridization 
$t_{pp}^{jk} = \pm t_{pp}$ with the minus sign occurring for 
$k = j - \hat{x}/2 - \hat{y}/2$ and $k = j + \hat{x}/2 + \hat{y}/2$, 
and the O-site energy $\epsilon_p$. The charge-transfer 
energy is defined as $\epsilon=\epsilon_{p}-\epsilon_{d}$.
For simplicity, we introduce dimensionless
parameters: $u=\alpha |u_{j}|/t_{pd}$, or $u=\beta |u_{j}|/t_{pd}$.
In units of $t_{pd}$, we use the set of parameters: $U_{d}=6$, 
$\epsilon=3$, $U_{p}=V_{pd}=0$, and $t_{pp}=0.5$.

All the results reported here were done on a lattice with $6 \times 6$ 
unit cells and periodic boundary conditions. In the following we will 
focus on the half-breathing O$(\pi,0)$ phonon mode and examine the 
dependence of physical quantities on the lattice displacement $u$
in different cases. In order to diminish systematic and statistical errors 
in the CPMC simulations, we studied two closed-shell cases:
the hole doping $\delta_{h}=\frac{6}{36}$ for the hole-doped case and 
the electron doping $\delta_{e}=\frac{10}{36}$ for the electron-doped case. 
Notice that we were forced to choose two different dopings for 
the hole- and electron-doped cases, due to the requirement of 
a closed-shell condition. However, we have checked that the results
at a smaller electron doping $\delta_{e}=\frac{10}{64}$ on a lattice with 
$8 \times 8$ unit cells remain qualitatively similar to the ones at
$\delta_{e}=\frac{10}{36}$, demonstrating that our findings on the $6\times 6$
lattice reflect intrinsic properties of the three-band Peierls-Hubbard model.

\begin{center}
\begin{figure}
\epsfig{file=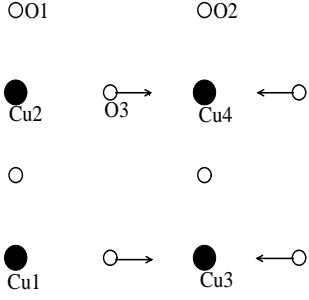,height=4.0cm,width=4.0cm,angle=270}
\smallskip
\caption{Displacement pattern of the half-breathing O$(\pi,0)$ mode
on the Cu-O plane.
The closed circles label the Cu sites, and the open circles
the O sites. The arrows denote the lattice displacements 
from the O sites.}
\label{LattB3}
\end{figure}
\end{center}

Fig.~\ref{LattB3} shows the displacement pattern of 
the half-breathing O$(\pi,0)$ phonon mode. Cu1, Cu2, Cu3 
and Cu4 label the copper sites, and O1, O2 and O3 
different oxygen sites. For the covalent el-ph coupling, 
the hybridization term $t_{pd}^{ij}$ is modified ($\alpha\neq 0$ and 
$\beta=0$), whereas  for the ionic el-ph coupling, the Cu-site energy 
$\epsilon_{d}^{i}$ is changed  ($\alpha=0$ and $\beta\neq 0$).

To clarify different effects of lattice displacement, 
we examine the $u$ dependence of\\
(i) the electronic energy 
\begin{equation}
E=\langle H \rangle,
\end{equation}
(ii) the kinetic energy
\begin{eqnarray}
E_{K}&=&\langle\sum_{\langle i,j\rangle\sigma}t_{pd}^{ij}(\{u_{j}\})
(d_{i\sigma}^{\dagger}p_{j\sigma}+h.c.)+\nonumber\\
&&\sum_{\langle j,k\rangle\sigma}
t_{pp}^{jk}(p_{j\sigma}^{\dagger}p_{k\sigma}+h.c.)\rangle,
\end{eqnarray}
(iii) the Cu-site potential energy
\begin{equation}
E_{P}=\langle\sum_{i\sigma}\beta(\sum_{j}\pm u_{j})n_{i\sigma}^{d}\rangle,
\end{equation}
(iv) the O-site energy
\begin{equation}
E_{O}=\langle\epsilon_{p}\sum_{j\sigma}n_{j\sigma}^{p}\rangle,
\end{equation}
(v) and the Coulomb energy 
\begin{equation}
E_{C}=\langle U_{d}\sum_{i}n_{i\uparrow}^{d}
n_{i\downarrow}^{d}\rangle,
\end{equation}
Here, $\langle\cdots\rangle$ means the ground-state expectation value.

\begin{center}
\begin{figure}
\epsfig{file=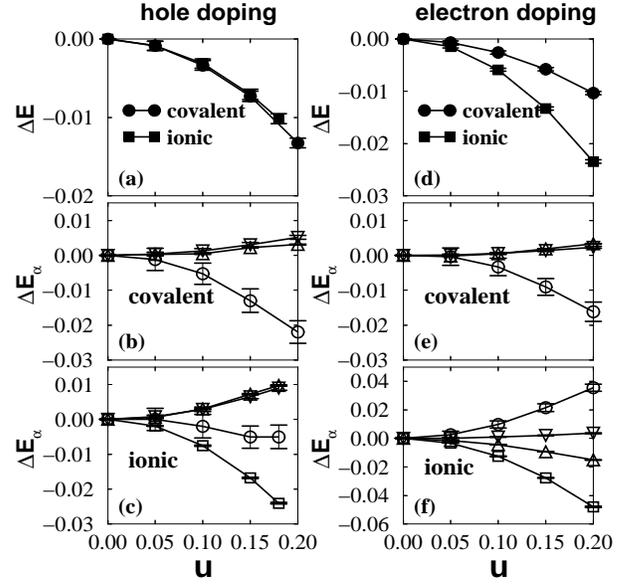,height=8cm,width=8cm,angle=270}
\smallskip
\caption{(a) Change $\Delta E$ (per unit cell) in electronic energy 
vs the lattice displacement $u$ in the hole-doped case; 
(b) changes $\Delta E_{\alpha}$ (per unit cell) in kinetic energy 
(open circles), O-site energy (open up-triangles), and Coulomb energy
(open down-triangles) vs the lattice displacement $u$ for the covalent
el-ph coupling in the hole-doped case;
(c) same as (b) for the ionic el-ph coupling,  and open squares represent 
the Cu-site potential energy; (d), (e), and (f) same as 
(a), (b), and (c), respectively, in the electron-doped case.}
\label{Energy}
\end{figure}
\end{center}

In Figs.~\ref{Energy}a, \ref{Energy}b, and \ref{Energy}c,  we show the changes
of different energies as a function of the lattice 
displacement $u$ in the hole-doped case.
The corresponding results in the electron-doped case are shown in 
Figs.~\ref{Energy}d, \ref{Energy}e, and \ref{Energy}f. 
In both hole- and electron-doped cases, the electronic energy is reduced by 
lattice displacements. As one can see, in the hole-doped case, the covalent 
and ionic el-ph couplings have an equal effect in the reduction of electronic 
energy, while in the electron-doped case, the ionic el-ph coupling 
has a stronger effect than the covalent el-ph coupling.

From Figs.~\ref{Energy}b and \ref{Energy}e, for the covalent el-ph coupling, 
one can see that the reduction of electronic energy comes from the gain 
in kinetic energy, while for the ionic el-ph coupling it originates from the gain
in Cu-site potential energy as shown in Figs.~\ref{Energy}c and \ref{Energy}f. 
Notice that although the noninteracting model ($U_{d}=0$) also decreases 
its electronic energy by lattice displacements through the same
mechanism, i.e., the gain in kinetic energy or Cu-site potential energy,
electronic correlations play an important role in the strongly correlated system. 
As a matter of facts, for a given lattice displacement $u$ 
the Coulomb interaction $U_{d}$ transfers some charges 
from copper sites to oxygen sites, making the kinetic energy $E_{K}$ 
in the interacting system even lower than that in the noninteracting system. 
This demonstrates that electronic correlations contribute essentially
to the gain in kinetic energy.
In addition, we observe that for the ionic el-ph coupling 
and in the noninteracting model, the Cu-site potential energy 
decreases more in the hole-doped case than in the electron-doped case. 
On the other hand, Figs.~\ref{Energy}c and 
\ref{Energy}f show that in the strongly correlated system this behavior 
is reversed, which is consistent with the physical situation where the doped
electrons mainly go to the copper sites, as a result, the charge fluctuations
at the copper sites are much stronger than those in the hole-doped case.

\begin{center}
\begin{figure}
\epsfig{file=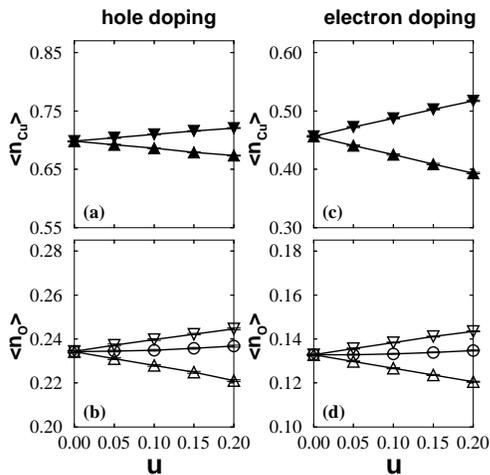,height=6.5cm,width=6.5cm,angle=270}
\smallskip
\caption{(a) Charge densities at the copper sites Cu1 (filled up-triangles)
and Cu3 (filled down-triangles) vs $u$ for the covalent el-ph coupling
in the hole-doped case; (b) same as (a) but at the oxygen sites O1
(open up-triangles), O2 (open down-triangles), and O3 (open circles);
(c) and (d) same as (a) and (b), respectively, in the electron-doped case.}
\label{Charge}
\end{figure}
\end{center}

In general, the strong el-ph coupling in both p- and n- type cuprates can be 
attributed to the strong interactions of lattice displacements with local metallic 
charge fluctuations. McQueeney {\it et al.}~\cite{queeney} have studied
the doping dependence of phonon densities of states in La$_{2-x}$Sr$_x$
CuO$_4$ and found that the development of the 70 meV phonon mode
near the doping-induced metal-insulator transition is due to the low-energy
one-dimensional charge dynamics. 
Moreover, Ishihara {\it et al.}~\cite{ishihara1,ishihara2} have studied
the one-dimensional two-band Peierls-Hubbard model and  indicated that
the charge-transfer fluctuations are strongly enhanced near the transition
from a Mott insulator to an ionic insulator (or a metal). As shown in
Fig.~\ref{Charge}, due to lattice displacements, the charge densities at
both copper and oxygen sites are modified in a similar way for 
the hole- and electron-doped systems. However, in our cases the
charge-transfer fluctuations have little contribution to the covalent
el-ph coupling, which is evidenced by a rather small change of
the O-site energy (see Figs.~\ref{Energy}b and \ref{Energy}e).

Now we turn to discuss the effects of lattice displacements on the
magnetic properties. Previous studies~\cite{schrieffer,eder}
have suggested that the quasiparticles of the low-energy excitations
in an antiferromagnet can be well described by the doped hole or electron 
dressed by local antiferromagnetic spin fluctuations (spin-bag picture). 
Therefore, it is important to understand the effects of lattice displacements
on local spin correlations in order to 
shed light on the single-particle excitation as measured in ARPES. 
Here, we are interested in the spin correlations of the next-nearest-neighbor 
(nnn) Cu-Cu sites, which are defined as
\begin{equation}
\label{spin}
S(i,j)=\langle{\bf S_{i} \cdot S_{j}}\rangle,
\end{equation}
with the spin operator ${\bf S_{i}}=
\sum_{\alpha\beta}d_{i\alpha}^{\dagger}{\bf\sigma_{\alpha\beta}}d_{i\beta}$.

\begin{center}
\begin{figure}
\epsfig{file=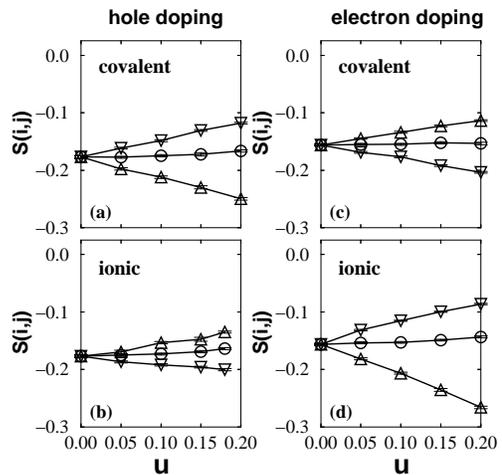,height=6.5cm,width=6.5cm,angle=270}
\smallskip
\caption{Spin correlations between the nnn Cu sites 
vs $u$ for the covalent el-ph coupling 
in the hole-doped case; 
(b) same as (a) for the ionic el-ph coupling; 
(c) and (d) same as (a) and (b), respectively,
in the electron-doped case.
The open circles, open up- and down-triangles represent the site
pairs (i,j)=(Cu1, Cu3), (i,j)=(Cu1, Cu2), and (i,j)=(Cu3, Cu4) 
(see Fig.~\ref{LattB3}), respectively.
}
\label{SpinCu_Cu}
\end{figure}
\end{center}

Figure~\ref{SpinCu_Cu} shows the nnn Cu-Cu spin correlations
as a function of $u$. One can clearly see that for both hole- and
electron-doped cases, the el-ph couplings have significant 
influences on the nnn Cu-Cu spin correlations:
while the spin correlations in the direction parallel to the
lattice displacement have little change, the spin correlations
in the direction perpendicular to the lattice displacement increase
or decrease alternatively depending on both the Cu sites and the 
doping. In particular, there exists a crucial difference between 
the hole- and electron-doped cases.
In the hole-doped case, the spin correlations increase 
for the Cu sites with smaller charge density, and decrease 
for the Cu sites with larger charge density, while
in the electron-doped case, this behavior is reversed.

\begin{center}
\begin{figure}
\epsfig{file=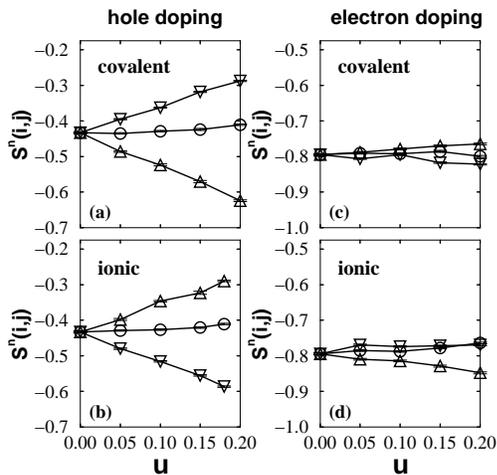,height=6.5cm,width=6.5cm,angle=270}
\smallskip
\caption{Same as Fig.~\ref{SpinCu_Cu} for the normalized spin
correlations.}
\label{SpinCu_m2}
\end{figure}
\end{center}

In order to filter out of the effects of charge fluctuations,
we have divided the spin operator by the probability of single
occupancy. The corresponding normalized spin correlations are 
defined as
\begin{equation}
\label{spincf}
S^{n}(i,j)={\langle{\bf S_{i} \cdot S_{j}}\rangle\over
P_{si}\cdot P_{sj}}.
\end{equation}
where $P_{si}=<n_{i}-2n_{i\uparrow}\cdot n_{i\downarrow}>$ denotes
the probability of single occupancy at the site $i$. Eq.~\eqref{spincf}, 
thus measures the spin correlations normalized to the spin density.
Figure~\ref{SpinCu_m2} shows the normalized spin correlations
as a function of $u$. From Figs.~\ref{SpinCu_m2}a and 
\ref{SpinCu_m2}b, it is clear that in the hole-doped case,
both ionic and covalent el-ph couplings induce a strong response
of spin degree of freedom to phonons. On the other hand, we observe
a rather weak effect in the electron-doped case, which is shown in
Figs.~\ref{SpinCu_m2}c and \ref{SpinCu_m2}d.

The above results indicate that there exist dramatic differences of
the spin-phonon coupling in p- and n-type high-T$_c$ cuprates:
the coupling of spin with lattice displacements is much stronger in the 
p-type materials than in the n-type materials. 
This could give an explanation why ARPES~\cite{shen} detected 
a strong mass renormalization in the p-type cuprates, while it is
absent in the n-type cuprates. The reason is that within the 
spin-bag picture, the quasiparticle property is mainly determined 
by local antiferromagetic spin fluctuations and a strong
change of local spin correlations can dramatically affect the motion
of doped charge carriers in the background of antiferromagnet.
We argue that if the similar difference in spin-phonon couplings 
occurs for other phonon modes at small and large momenta, 
we then expect that in the n-type superconductors, the phonon 
contributions to electron-electron pairing and electronic transport 
should be much smaller than those in the p-type materials.

Summarizing, we have presented QMC studies of the effects of 
the half-breathing O$(\pi,0)$ phonon mode on the ground-state
properties of the 2D three-band Peierls-Hubbard model.
We found that in the physically interesting parameter regime, 
the el-ph couplings have similar effects on the ground-state
energies and charge fluctuations in both hole- and electron-doped
systems. This suggests that the strong softening of the half-breathing 
O$(\pi,0)$ phonon mode in both p- and n-types materials can be
attributed to the the strong interactions of lattice distortions with 
local metallic charge fluctuations. In addition, our results show that 
the spin-phonon coupling is much stronger in the p-type cuprates 
than in the n-type cuprates. This difference in the spin-phonon 
coupling could provide an explanation for the difference of 
mass renormalization observed in ARPES experiments.

This work was supported by the DFG under Grant No.~Ha 1537/16-2 
and by a  Heisenberg Grant (AR 324/3-1), by the Bavaria California 
Technology Center (BaCaTeC), the  KONWHIR projects OOPCV and CUHE.
The calculations were carried out at the high-performance computing centers
HLRS (Stuttgart).

\end{document}